\documentclass[aps,prl,showpacs,twocolumn,superscriptaddress]{revtex4}
\usepackage{bm,color}
\usepackage{graphicx}
\usepackage{amsmath}
\usepackage{color}
\pdfoutput=1

\begin{document}
\title {Highly Efficient Induction of Spin Polarization by Circularly-Polarized Electromagnetic Waves in the Rashba Spin-Orbit Systems}

\author{Masahito Mochizuki}
\affiliation{Department of Applied Physics, Waseda University, Okubo, Shinjuku-ku, Tokyo 169-8555, Japan}
\affiliation{Department of Physics and Mathematics, Aoyama Gakuin University, Sagamihara, Kanagawa 229-8558, Japan}
\affiliation{PRESTO, Japan Science and Technology Agency, Kawaguchi, Saitama 332-0012, Japan}
\author{Keisuke Ihara}
\affiliation{Department of Physics and Mathematics, Aoyama Gakuin University, Sagamihara, Kanagawa 229-8558, Japan}
\author{Jun-ichiro Ohe}
\affiliation{Department of Physics, Toho University, 2-2-1, Miyama, Funabashi, Chiba 274-8510, Japan}
\author{Akihito Takeuchi}
\affiliation{Department of Physics and Mathematics, Aoyama Gakuin University, Sagamihara, Kanagawa 229-8558, Japan}
\begin{abstract}
We theoretically demonstrate that a rotating {\it electric-field} component of circularly polarized microwave or terahertz light can induce electron-spin polarization within a few picoseconds in a two-dimensional electron system with the Rashba spin-orbit interaction by taking advantage of the magnetoelectric coupling. The efficiency turns out to be several orders of magnitude greater than that of conventional methods, indicating high potential of this technique for future spintronics.
\end{abstract}
\maketitle

The fast and efficient manipulation of electron spins in magnetic semiconductors is a vital issue in recent spintronics research~\cite{Maekawa06,Wolf01,Zutic04,Chappert07,ScurrentBook}. Induction, switching and driving of magnetization can be achieved by application of external stimuli such as magnetic fields~\cite{Back98,Back99,Kaka02,Gerrits02,Schumacher03,Tudosa04} and electric currents~\cite{Slonczewski96,Berger96,Parkin08,Yamanouchi04}. These phenomena are expected to be useful for technical applications in magnetic devices. The techniques based on the application of magnetic field exploit the Zeeman coupling between the electron spins and the magnetic field, whereas those based on the electric-current injection take advantage of, for example, angular-momentum transfer from conduction-electron spins to the local magnetization~\cite{Bazaliy98,Tatara04,Barnes05,Brataas12}.

\begin{figure}
\begin{center}
\includegraphics[width=1.0\columnwidth]{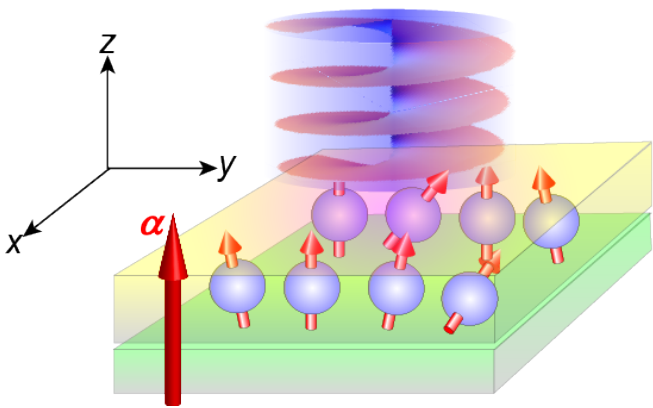}
\caption{(color online). (a) Schematic illustration of the magnetzation induction by irradiating a circularly polarized microwave or terahertz light in the two-dimensional electron system with broken spatial inversion symmetry where the Rashba spin-orbit interaction is active.}
\label{Fig1}
\end{center}
\end{figure}
The Rashba spin-orbit interaction (SOI) provides another important physical mechanism to manipulate magnetism by electric current injection~\cite{Rashba60,Vasko79,Bychkov84}, and it has been attracting a great deal of interest because of its strength and controllability via application of a gate voltage~\cite{Nitta97,Engels97,Schultz96}. This interaction becomes prominent in asymmetric heterostructures of narrow-gap semiconductors. A number of interesting spin-dynamics and spin-transport phenomena related to the Rashba interaction have been theoretically proposed and experimentally demonstrated~\cite{Manchon15,Kohda08,Jungwirth12}: the spin-galvanic effects (the spin-to-charge current conversion)~\cite{Ivchenko78,Ganichev02,Ganichev08,Wunderlich09,Rojas13,Ohe07}; the inverse spin-galvanic effects (the charge current-to-spin conversions)~\cite{Edelstein90,Kato04,Ganichev06}; spin-polarized field-effect transistors~\cite{Datta90,KooHC09}, spin-current generation via the spin filtering effects or spin-polarizing effects~\cite{Kiselev01,Pareek04,Koga02,Ionicioiu03,Nitta99,Ohe05,Kohda12,Debray09}, and the electrical manipulation of spins~\cite{Nitta09,Kohda12b,Kunihashi16}

However, techniques based on the electric current injection encounter a serious problem of large energy consumption due to Joule-heating losses; even techniques based on the magnetic field application cannot avoid it because the magnetic field is generated via application of electric currents to a metallic coil. Conversely, techniques using microwaves and terahertz (THz) light have high potential to realize significant reduction of energy losses. Recently, induction and switching of magnetization with a circularly polarized electromagnetic waves have been proposed theoretically~\cite{Takayoshi14a,Takayoshi14b}. The physical mechanism of this effect is a direct activation of the electron-spin precession through coupling between the circularly polarized {\it magnetic field} and the electron spins via the conventional Zeeman interaction. Its efficiency is, however, limited because an energy of the Zeeman interaction is very small. A relation $B_0=E_0/c$ holds between the amplitude of alternating current (AC) magnetic field $B_0$ and that of AC electric field $E_0$ for electromagnetic waves. This means that $B_0$ is only $\sim$0.3 T even for a relatively intense laser with $E_0$=1 MV/cm, which corresponds to a Zeeman energy of only 18 $\mu$eV. Therefore, an enhancement of the effect has been called for.

In this Letter, we theoretically propose that tremendous enhancement of the effect can be realized at microwave or THz frequencies by exploiting the {\it electric-field} component of the electromagnetic wave. It requires irradiation of a circularly polarized laser or microwave onto a two-dimensional electron system with the Rashba SOI [see Fig.~\ref{Fig1}]. Its rotating electric-field component generates an effective rotating magnetic field via the Rashba SOI, and its amplitude is several orders of magnitude higher than that of the original AC magnetic-field component, e.g., four orders of magnitude larger at 10 THz. The present method can enable manipulation of electron spins with much greater efficiency than previously proposed methods using the circularly polarized magnetic-field component, and thus will advance research of the spintronics research onto a next stage.

We study the effects of circularly polarized light irrdiation on a two-dimensional electron gas with the Rashba SOI realized at the interface of asymmetric heterostructure of semiconductors. By numerically solving the time-dependent Schr\"odinger equation, we trace time evolutions of the Gaussian wave packets with up and down spins, $\vec{\psi}_{\uparrow}(\bm r, t)$ and $\vec{\psi}_{\downarrow}(\bm r, t)$.

The time-dependent Schr\"odinger equation reads
\begin{eqnarray}
i\hbar\frac{\partial \vec{\psi}}{\partial t}=\mathcal{H} \vec{\psi}.
\label{eq:Schrodinger}
\end{eqnarray}
The Hamiltonian $\mathcal{H}$ is composed of four terms as,\\
$\mathcal{H}=\mathcal{H}_{\rm K}+\mathcal{H}_{\rm R}+\mathcal{H}_E+\mathcal{H}_B$ where
\begin{eqnarray}
\mathcal{H}_{\rm K}&=&-\frac{\hbar^2}{2m}\left(\frac{\partial^2}{\partial x^2}
+\frac{\partial^2}{\partial y^2} \right),
\label{eq:kinetic} \\
\mathcal{H}_{\rm R}&=&\frac{1}{\hbar}\bm \sigma \cdot
\left( \bm p \times \bm \alpha \right)
=-\frac{\alpha}{\hbar}\left(p_x \sigma_y - p_y \sigma_x \right),
\label{eq:Rashba} \\
\mathcal{H}_E&=&-e\phi_{\rm em}
=e E_0 \left(x E_x^\omega(t) + y E_y^\omega \right),
\label{eq:elepot} \\
\mathcal{H}_B&=&\frac{g\mu_{\rm B}}{2} 
\left(\sigma_x B_x^\omega(t) + \sigma_y B_y^\omega(t) \right).
\label{eq:zeeman}
\end{eqnarray}
Here $e(>0)$ is the elementary charge, and $m$ is the electron mass. We assume a free-electron mass for $m$ in this study.
The first term $\mathcal{H}_{\rm K}$ represents the kinetic energy. The second term $\mathcal{H}_{\rm R}$ represents the Rashba interaction where $\bm \alpha=\alpha \bm e_z$ is the Rashba vector with $\alpha$ being the Rashba parameter. We take $\alpha$=0.6 eV$\cdot$m as a typical value of the Rashba parameter~\cite{Nitta97,Engels97}. 
The third term $\mathcal{H}_E$ describes the AC electrical potential induced by the rotating electric field $\bm E^\omega(t)$ of the circularly polarized light. The fourth term $\mathcal{H}_B$ describes the Zeeman coupling between the electron spin and the rotating magnetic field $\bm B^\omega(t)$ of the light. The $\bm E^\omega(t)$ and $\bm B^\omega(t)$ fields are given, respectively, by
\begin{align}
&\bm E^\omega(t)=E_0\beta(t)
\left(\cos\omega t, \sin(\omega t+\varphi) \right),
\\
&\bm B^\omega(t)=\frac{E_0}{c}\beta(t)
\left(\sin(\omega t+\varphi),-\cos\omega t \right).
\end{align}
Here the time-dependent factor $\beta(t)=\tanh(t/\tau_1)$ with $\tau_1$=0.868 ps is introduced for a gradual rise of the external fields to observe essential phenomena induced by the Rashba SOI avoiding intense and nonlinear spin oscillations in the initial process caused by an impact of abruptly irradiated electromagnetic waves. The value of $\varphi$ determines the helicity of the laser, i.e., $\varphi$=0, $\pi/2$, and $\pi$ correspond to left-circularly, linearly and right-circularly polarized lights, respectively.

The initial forms of the wave packets are given by,
\begin{align}
\vec{\psi}_{\sigma}(t=0)
&=\frac{1}{\sqrt[4]{2\pi\delta^2}}\exp\left[-\frac{x^2+y^2}
{4\delta^2}\right]\vec{\chi}_{\sigma}
\label{eq:Gaussian}
\end{align}
with $\delta$=6.75 nm. The quantization axis of spin is chosen to be the $z$ axis, and the spin-parts of the wavefunctions are given by,
\begin{equation}
\vec{\chi}_{\uparrow}=\begin{pmatrix}1 \\ 0\end{pmatrix}, \quad
\vec{\chi}_{\downarrow}=\begin{pmatrix}0 \\ 1\end{pmatrix}.
\end{equation}

For the numerical simulations, we employ a system of $L_x \times L_y$ =100 {\rm nm} $\times$ 100 {\rm nm} with an open boundary condition. Dividing this system into idential square cells of $a \times a$ with $a$=2 nm, the Laplacian in the Schr\"odinger equation is treated within the finite-difference method. We solve this differential equation using the fourth-order Runge-Kutta method numerically. Spatial distributions of the dynamical spin density $D_{\rm s}$ and the dynamical charge density $D_{\rm c}$ are calculated using the simulated wavefunctions $\psi_{\uparrow}(\bm r, t)$ and $\psi_{\downarrow}(\bm r, t)$ as,
\begin{eqnarray}
D_{\rm s}(\bm r, t)&=&\sum_{\sigma} 
 |\vec{\chi}_{\uparrow}   \cdot \vec{\psi}_{\sigma}(\bm r, t)|^2
-|\vec{\chi}_{\downarrow} \cdot \vec{\psi}_{\sigma}(\bm r, t)|^2, \\
D_{\rm c}(\bm r, t)&=&\sum_{\sigma} 
 |\vec{\chi}_{\uparrow}   \cdot \vec{\psi}_{\sigma}(\bm r, t)|^2
+|\vec{\chi}_{\downarrow} \cdot \vec{\psi}_{\sigma}(\bm r, t)|^2,
\end{eqnarray}
where $\sigma$=$\uparrow$ and  $\downarrow$.

\begin{figure}
\includegraphics[width=1.0\columnwidth]{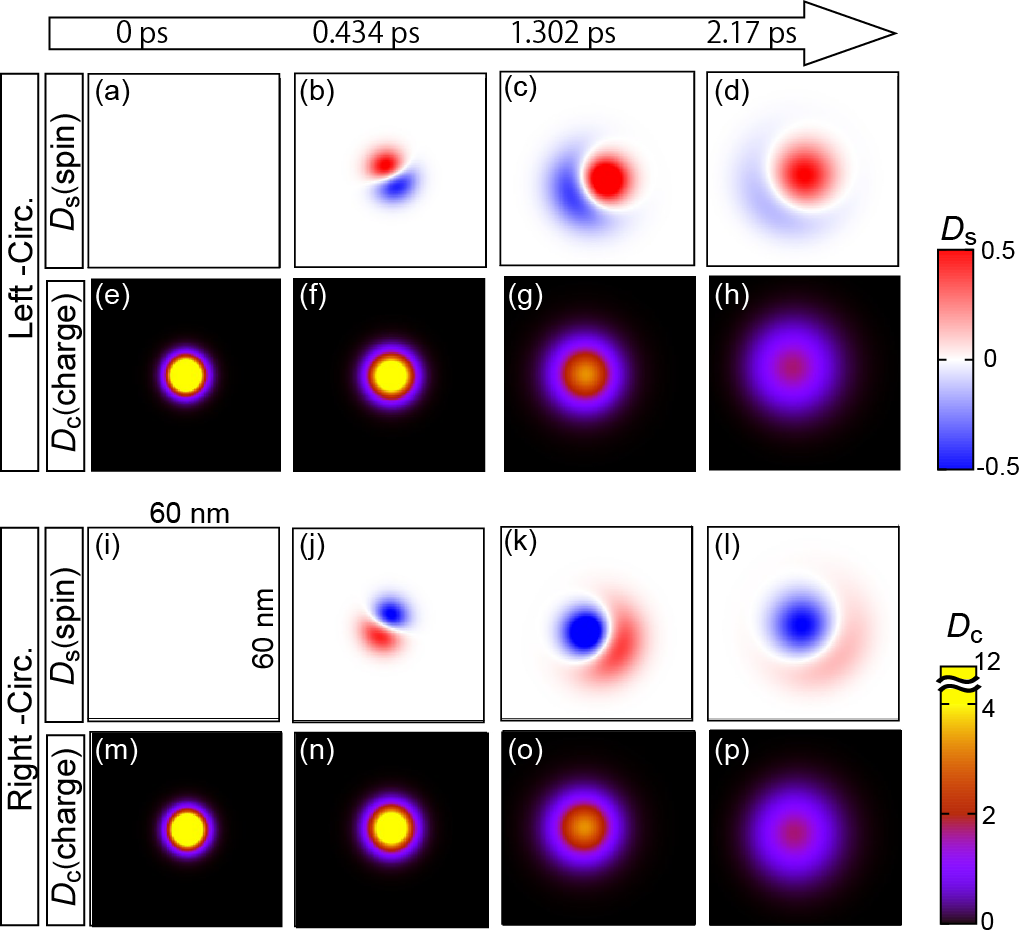}
\caption{(color). Snapshots of the simulated spin and charge density distributions ($D_{\rm s}$ and $D_{\rm c}$) for selected times when left-circularly or right-circularly polarized light is used. (a)-(d) [(e)-(h)] Spin [Charge] density distribution under a left-circularly polarized light. (i)-(l) [(m)-(p)] Spin [Charge] density distribution under a right-circularly polarized light. The relevant area of 60 nm $\times$ 60 nm is magnified, while the simulations are performed for a larger system size of 100 nm $\times$ 100 nm.}
\label{Fig2}
\end{figure}
In Fig.~\ref{Fig2}, we display snapshots of the simulated spin density $D_{\rm s}$ and charge density $D_{\rm s}$ at selected times under irradiation of left-circularly polarized and right-circularly polarized light. Here the amplitude and the frequency of the light are fixed at $E_0$=1 MV/cm and $\omega/2\pi$=10 THz, respectively. Initially, the spin density is zero in total because of the cancellation between equally weighted up-spin and down-spin wave packets, $\psi_{\uparrow}(\bm r, 0)$ and $\psi_{\downarrow}(\bm r, 0)$. Immediately after starting the laser irradiation, the Gaussian wave packet splits into two portions with opposite spin components, i.e., a portion with a dominant up-spin density and one with a dominant down-spin density. The resulting spin-density dipole circulates in the same direction as the rotations of the light $\bm B^\omega(t)$ and $\bm E^\omega(t)$ fields. Through this spin-polarizing process, the charge-density distribution gradually spreads because of the diffusion of the Gaussian wave packets. In a series of simulated charge-density snapshots with a shorter interval (not shown), we find that the Gaussian wave packet circulates in a counterclockwise (clockwise) fashion for the left-circularly (right-circularly) polarized case in accord with the rotation direction of the $\bm E^\omega$ and $\bm B^\omega$ components.

\begin{figure}
\includegraphics[width=1.0\columnwidth]{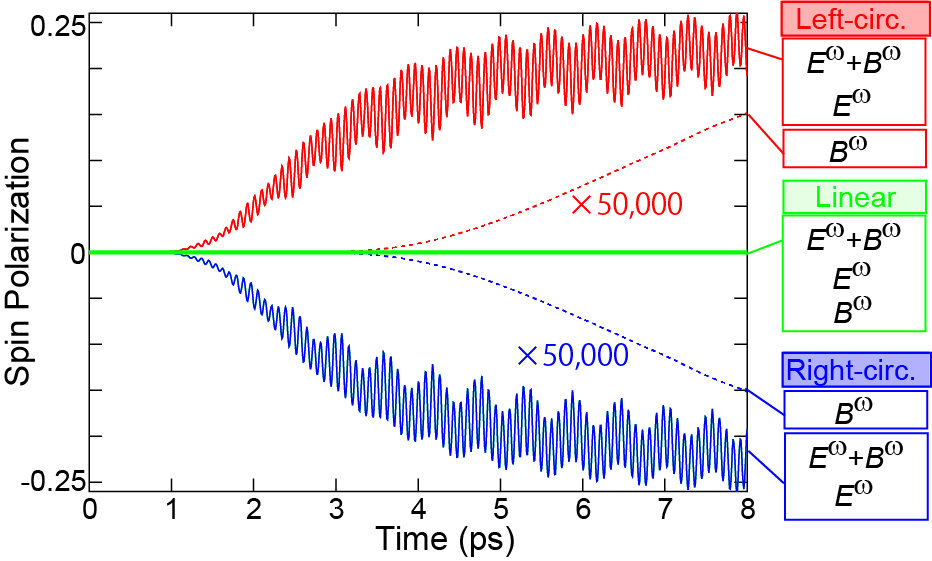}
\caption{(color online). Simulated time evolution of expectation values of the spin $z$-axis component $\bar{S}(t)$ under left-circularly, right-circularly and linearly polarized lights with $\bm E^\omega + \bm B^\omega$ fields. We also plot the results for the case with the electric-field component $\bm E^\omega$ only and the case with the magnetic-field component $\bm B^\omega$ only.}
\label{Fig3}
\end{figure}
In Fig.~\ref{Fig3}, we plot time profiles of the expectation value of the spin $z$-axis component calculated as
\begin{eqnarray}
\bar{S}(t)=\int d\bm r D_{\rm s}(\bm r, t)/\int d\bm r D_{\rm c}(\bm r, t)
\end{eqnarray}
for different light polarizations, i.e., left-circularly, right-circularly, and linearly polarized lights with both $\bm E^\omega(t)$ and $\bm B^\omega(t)$ fields (indicated as $\bm E^\omega + \bm B^\omega$ in the figure). Here the amplitude and the frequency of the circularly polarized lights are fixed at $E_0$=1 MV/cm and $\omega/2\pi$=10 THz, respectively. We find that $\bar{S}(t)$ grows very quickly (a typical time scale is a few picoseconds) for circularly polarized lights. Importantly, the sign of the induced spin component differs depending on the handedness of the circular light-polarization, that is, it is positive (negative) for a left-circularly (right-circularly) polarized case. In contrast, the expectation value remains zero for the linearly polarized case, indicating that the magnetization cannot be induced by the linearly polarized light.

Simulations are performed also for the case with the electric-field component ($\bm E^\omega$) only and the case with the magnetic-field component ($\bm B^\omega$) only. The results obtained for $\bm E^\omega$ only show almost perfect coincidence with those for $\bm E^\omega + \bm B^\omega$ in the plot. However, the results for $\bm B^\omega$ only are negligibly small, so that we need to multiply by 50,000 to make them visible in the present plot scale. We should also mention that the Gaussian wave packets show diffusive behaviors because they are not eigenstates of the Hamiltonian, which phenomenologically mimics the effects of scattering and relaxation of the electron spin in real systems, resulting in the incomplete spin polarization seen in Fig.~\ref{Fig3}.

Now we discuss a physical mechanism for the observed enhanced spin-polarization induction by irradiation of a circularly polarized electromagnetic wave in the Rashba SOI system. By comparing Eq.~(\ref{eq:Rashba}) with a formula of the usual Zeeman interaction
\begin{align}
\mathcal{H}_{\rm Zeeman}=g \mu_{\rm B} \bm B \cdot 
\frac{1}{2}\bm \sigma,
\end{align}
we find that the Rashba SOI induces an effective magnetic field
\begin{align}
\bm B_{\rm eff}=\frac{2\alpha}{g \mu_{\rm B}\hbar} 
\left(\bm p \times \bm e_z \right).
\end{align}
This effective magnetic field $\bm B_{\rm eff}$ acts on moving conduction-electron spins, and its magnitude and direction are governed by a momentum $\bm p$ of the electron as well as the Rashba vector $\bm \alpha(=\alpha \bm e_z)$.

\begin{figure}
\includegraphics[width=1.0\columnwidth]{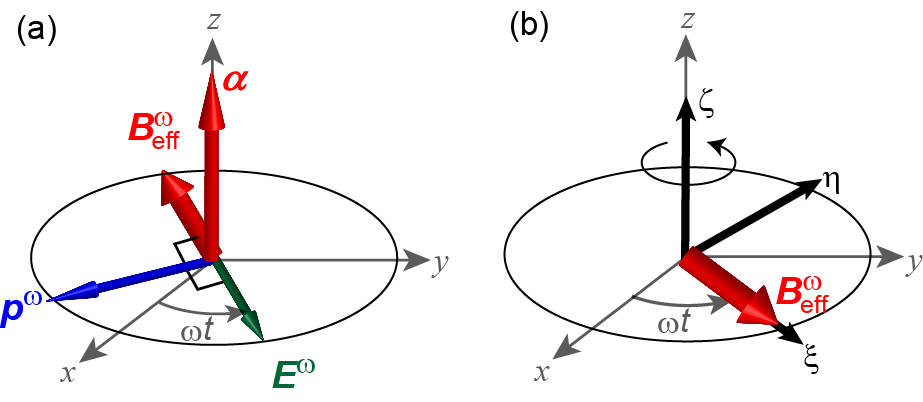}
\caption{(color online). (a) Rotating electric-field component $\bm E^\omega$ of the circularly polarized light induces a momentum $\bm p^\omega$ of an electron, which leads to an effective rotating magnetic field $\bm B_{\rm eff}^\omega$ ($\parallel \bm p \times \bm \alpha$) in the presence of the Rashba SOI with the Rashba vector $\bm \alpha$ normal to the two-dimensional plane. (b) Induced rotating effective in-plane magnetic field $\bm B_{\rm eff}^\omega$ leads to a fictitious rotating magnetic field with an out-of-plane component $-\omega/\gamma$ with respect to the coordinates rotating with the $\bm B_{\rm eff}^\omega$ field.}
\label{Fig4}
\end{figure}

Accordingly, the rotating electric field $\bm E^\omega$ of circularly polarized light can induce a rotating effective magnetic field $\bm B_{\rm eff}^\omega$ in the presence of the Rashba SOI, which causes the electron-spin polarization as formulated in the following. When an electron moves under the influence of the rotating electric field
$\bm E^\omega=E_0 \left(\cos\omega t, \sin\omega t\right)$,
an equation of motion for the electron is given by,
\begin{align}
m\begin{pmatrix}\ddot{x} \\ \ddot{y}\end{pmatrix}=
-eE_0\begin{pmatrix}\cos\omega t \\ \sin\omega t\end{pmatrix}
-\frac{m}{\tau}\begin{pmatrix}\dot{x} \\ \dot{y}\end{pmatrix}
\end{align}
where $\tau$ is the relaxation rate. In the high-frequency limit ($\omega\tau \gg 1$), this gives a momentum $\bm p^\omega$ of the electron as
\begin{align}
\bm p^\omega=m\begin{pmatrix}\dot{x} \\ \dot{y}\end{pmatrix}=
\frac{eE_0}{\omega}\begin{pmatrix}-\sin\omega t \\ \cos\omega t\end{pmatrix}.
\end{align}
Here we take $\bm p^\omega=0$ for the initial momentum at $t$=0 because the momentum averaged over all the electrons should vanish when the external field is absent. This momentum gives rise to an effective rotating magnetic field $\bm B_{\rm eff}^\omega$ via the Rashba SOI. The expression of $\bm B_{\rm eff}^\omega$ is given by,
\begin{align}
\bm B_{\rm eff}^\omega=-\frac{2e\alpha E_0}{g\mu_{\rm B}\hbar \omega}
\begin{pmatrix}\cos\omega t \\ \sin\omega t\end{pmatrix}.
\end{align}
The amplitude of this effective rotating magnetic field is proportional to both the Rashba parameter $\alpha$ and the electric-field intensity $E_0$, whereas it is inversely proportional to the angular frequency $\omega$. Note that this $\bm B_{\rm eff}^\omega$ field points in the opposite direction to the $\bm E^\omega$ field [see Fig.~\ref{Fig4}(a)].

When the Rashba parameter is $\alpha$=6.0 meV$\cdot$nm and the intensity of the circularly polarized light is $E_0$=1 MV/cm, the amplitude of the Rashba-induced $\bm B_{\rm eff}^\omega$ field reaches 4300 T for $\omega/2\pi$=10 THz, which is four orders of magnitude larger than the amplitude of the original AC magnetic field of the light, i.e., $B_0=E_0/c\sim0.3$ T.

Induction of the spin $z$-axis component by application of this effective rotating magnetic field in the $xy$ plane can be understood as follows. When the rotating magnetic field is represented as $\bm B_{\rm eff}^\omega=B_0^\omega(\cos \omega t, \sin \omega t, 0)$ with respect to the rest coordinates, the spin vector $\bm S$ is represented with respect to the coordinates rotating with this field as,
\begin{eqnarray}
\bm S=S_\xi \bm e_\xi + S_\eta \bm e_\eta + S_\zeta \bm e_\zeta,
\label{eq:deriv1}
\end{eqnarray}
where $\bm e_\xi=(\cos \omega t, \sin \omega t, 0)$, $\bm e_\eta=(-\sin \omega t, \cos \omega t, 0)$ and $\bm e_\zeta=(0, 0, 1)$ are unit directional vectors of the rotating coordinate system [see Fig.~\ref{Fig4}(b)]. The $\xi$-axis is chosen to be parallel to the $\bm B_{\rm eff}^\omega$ field.
Taking the time-derivative for both sides of Eq.~(\ref{eq:deriv1}), we obtain
\begin{eqnarray}
\left. 
\frac{d\bm S}{dt}=\frac{d\bm S}{dt} \right|_{\rm R}
+\omega \left(S_\xi \bm e_\eta - S_\eta \bm e_\xi \right),
\label{eq:deriv2}
\end{eqnarray}
where $( ... )_{\rm R}$ denotes a vector represented using $\bm e_x$, $\bm e_y$ and $\bm e_z$ as bases or a vector represented with respect to the rotating coordinates. Here we use the following relations in the derivation:
\begin{eqnarray}
& &
\left. 
\frac{d\bm S}{dt} \right|_{\rm R}
=\frac{dS_\xi}{dt} \bm e_\xi + \frac{dS_\eta}{dt} \bm e_\eta 
+ \frac{dS_\zeta}{dt} \bm e_\zeta, 
\\
& &
\frac{d\bm e_\xi}{dt}=\omega \bm e_\eta, 
\quad
\frac{d\bm e_\eta}{dt}=-\omega \bm e_\xi,
\quad
\frac{d\bm e_\zeta}{dt}=0.
\label{eq:deriv3}
\end{eqnarray}
Substituting the following two relations,
\begin{eqnarray}
\frac{d\bm S}{dt}=-\gamma \bm S \times \bm B=-\gamma
\begin{pmatrix}S_\xi \\ S_\eta \\ S_\zeta\end{pmatrix}_{\rm R}
\times
\begin{pmatrix}B_0^\omega \\ 0 \\ 0\end{pmatrix}_{\rm R}
\label{eq:deriv4}
\end{eqnarray}
and
\begin{eqnarray}
\omega \left(S_\xi \bm e_\eta - S_\eta \bm e_\xi \right)=
\omega \begin{pmatrix}-S_\eta \\ S_\xi \\ 0\end{pmatrix}_{\rm R}=
\begin{pmatrix}S_\xi \\ S_\eta \\ S_\zeta\end{pmatrix}_{\rm R}
\times
\begin{pmatrix}0 \\ 0 \\ -\omega\end{pmatrix}_{\rm R}
\label{eq:deriv5}
\end{eqnarray}
into Eq.~(\ref{eq:deriv3}), we obtain
\begin{eqnarray}
\begin{pmatrix}dS_\xi/dt \\ dS_\eta/dt \\ dS_\zeta/dt \end{pmatrix}_{\rm R}
=-\gamma
\begin{pmatrix}S_\xi \\ S_\eta \\ S_\zeta\end{pmatrix}_{\rm R}
\times
\begin{pmatrix}B_0^\omega \\ 0 \\ -\omega/\gamma\end{pmatrix}_{\rm R},
\label{eq:deriv6}
\end{eqnarray}
where $\gamma(=g\mu_{\rm B}/\hbar)$ is the electron gyromagnetic ratio. This equation indicates that the circularly-polarized in-plane magnetic field effectively works as a fictitious magnetic field given by $\bm B_{\rm R}=(B_0^\omega, 0, -\omega/\gamma)$, which contains a steady out-of-plane component proportional to the angular frequency $\omega$  and thus can induce a spin component perpendicular to the plane. In this way, the rotating effective magnetic field $\bm B_{\rm eff}^\omega$ generated by the circularly polarized electric field $\bm E^\omega$ via the Rashba SOI can induce or switch the electron spin magnetization.

In summary, we have theoretically demonstrated that irradiation of a circularly polarized microwave or THz light can efficiently induce electron-spin polarization in a two-dimensional electron system with the Rashba sSOI. The AC {\it electric field} of the circularly polarized electromagnetic wave generates an effective rotating magnetic field via the Rashba SOI, which can be several orders of magnitude larger in amplitude than its original AC magnetic-field component. Consequently, this phenomenon provides us with a highly efficient method to manipulate the electron spins, and will be a breakthrough for spintronics research. Since the Rashba SOI does not necessarily cause the spin relaxation for coherently driven electrons, the spin relaxation rate can be reduced in a cleaner electron gas system, in which more enhanced spin polarization can be expected with less spin damping effects. Development of intense THz light source is necessary for experimental feasibility of our proposal~\cite{Hirori11,Matsunaga13,Vicario13}. We assume a laser intensity of $E_0$$\sim$1 MV/cm, which is currently available only for a pulse with a few cycles. Experimental efforts to realize a continuous intense THz laser of circular polarization as well as efforts to realize a stronger Rashba coupling $\alpha$ will lead to an enhanced efficiency of the spin-polarization induction.

This research was supported in part by JSPS KAKENHI (Grant Nos. 25870169, 25287088, 17H02924, and 17H01052), Waseda University Grant for Special Research Projects (Project No. 2017S-101), and JST PRESTO (Grant No. JPMJPR132A).

\end{document}